\documentclass[letterpaper,english,reprint, aps]{revtex4-2}
\usepackage[T1]{fontenc}
\usepackage[utf8]{inputenc}
\usepackage{amsmath}
\usepackage{amssymb}
\usepackage{graphicx}
\usepackage{esint}

\makeatletter

\newcommand{\lyxmathsym}[1]{\ifmmode\begingroup\def\b@ld{bold}
  \text{\ifx\math@version\b@ld\bfseries\fi#1}\endgroup\else#1\fi}

\usepackage{bm}
\usepackage{float}
\usepackage{amsthm}
\usepackage{mathrsfs}
\usepackage{xcolor}
\usepackage{titlesec}

\usepackage{hyperref}
\hypersetup{
    colorlinks=true,            
    linkcolor=blue,              
    citecolor=blue,            
    filecolor=magenta,          
    urlcolor=blue               
}

\makeatother

\begin{document}
\preprint{APS/123-QED}
\title{Observation of topological Phenomena in a Weyl Exceptional Ring with
Single Photons}
\author{Zhong-Sheng Chen}
\thanks{These authors contributed equally to this work.}
\affiliation{Fujian Key Laboratory of Quantum Information and Quantum~~\\
 Optics, College of Physics and Information Engineering, Fuzhou University,
Fuzhou, Fujian, 350108, China}
\author{Wei-Xin Chen}
\thanks{These authors contributed equally to this work.}
\affiliation{Fujian Key Laboratory of Quantum Information and Quantum~~\\
 Optics, College of Physics and Information Engineering, Fuzhou University,
Fuzhou, Fujian, 350108, China}
\author{Fan Wu}
\email{T21060@Fzu.edu.cn}

\affiliation{Fujian Key Laboratory of Quantum Information and Quantum~~\\
 Optics, College of Physics and Information Engineering, Fuzhou University,
Fuzhou, Fujian, 350108, China}
\author{Zhong-Wei Xu}
\affiliation{Fujian Key Laboratory of Quantum Information and Quantum~~\\
 Optics, College of Physics and Information Engineering, Fuzhou University,
Fuzhou, Fujian, 350108, China}
\author{Jing Ma}
\affiliation{Fujian Key Laboratory of Quantum Information and Quantum~~\\
 Optics, College of Physics and Information Engineering, Fuzhou University,
Fuzhou, Fujian, 350108, China}

\author{Yun-Kun Jiang}
\email{ykjiang@fzu.edu.cn}

\affiliation{Fujian Key Laboratory of Quantum Information and Quantum~~\\
 Optics, College of Physics and Information Engineering, Fuzhou University,
Fuzhou, Fujian, 350108, China}

\author{Huai-Zhi Wu}
\email{huaizhi.wu@fzu.edu.cn}

\affiliation{Fujian Key Laboratory of Quantum Information and Quantum~~\\
	Optics, College of Physics and Information Engineering, Fuzhou University,
	Fuzhou, Fujian, 350108, China}

\author{Shi-Biao Zheng}

\affiliation{Fujian Key Laboratory of Quantum Information and Quantum~~\\
 Optics, College of Physics and Information Engineering, Fuzhou University,
Fuzhou, Fujian, 350108, China}
\date{\today}
\begin{abstract}
Compared with Hermitian theory, non-Hermitian physics offers a fundamentally
different mathematical framework, enabling the observation of topological
phenomena that have no analogue in Hermitian systems. Among these,
the exceptional point (EP) ring stands out as a quintessential topological
feature unique to non-Hermitian systems. In this study, we employ single-photon interferometry to overcome the experimental challenge of precise phase control in quantum systems, thereby enabling a complete simulation of the non-Hermitian EP ring in three-dimensional parameter space without invoking any additional symmetry assumptions. By measuring
the non-Hermitian dynamics in three-dimensional parameter space, we
determine the system's eigenstates, which allows us to characterize
the topological band structure of the system under different conditions.
We describe the topological properties of the EP ring by extracting
the Chern number and Berry phase for different parameter manifolds
and observe the topological critical phenomena of the system. Our
work paves the way for further exploration of topological non-Hermitian
systems. 
\end{abstract}
\maketitle

\section{Introduction}

In the past, most investigations of quantum mechanical phenomena have
been considered within the context of closed systems, aiming to minimize
decoherence arising from interaction with the environment. However,
the non-Hermitian effects induced by dissipation, once seen primarily
as a nuisance, have been shown, in certain regimes, to give rise
to novel features unattainable in conventional Hermitian systems \citep{1,2,3,4}.
These striking phenomena, such as the non-Hermitian skin effect and
anomalous bulk-boundary correspondence, are closely tied to the exceptional
points (EPs), where both eigenvalues and eigenvectors coalesce. Unlike
degeneracies in Hermitian systems, where eigenvalues may coincide
while eigenvectors remain orthogonal, EPs exhibit fundamentally different
behavior. The fascinating properties of EPs have not only spurred
a wide range of practical applications, including enhanced sensing \citep{5,6,7,8,9,10,11},
unidirectional wave propagation \citep{12,13,14,15}, chiral laser
emission \citep{16}, laser linewidth broadening \citep{17}, and laser
mode selection \citep{18,19,20,21}, but their underlying topological
structure has also drawn considerable attention in its own right \citep{22,23,24,25,26,29,30,31,32,33}.
\begin{figure*}
\includegraphics[width=1.3\columnwidth]{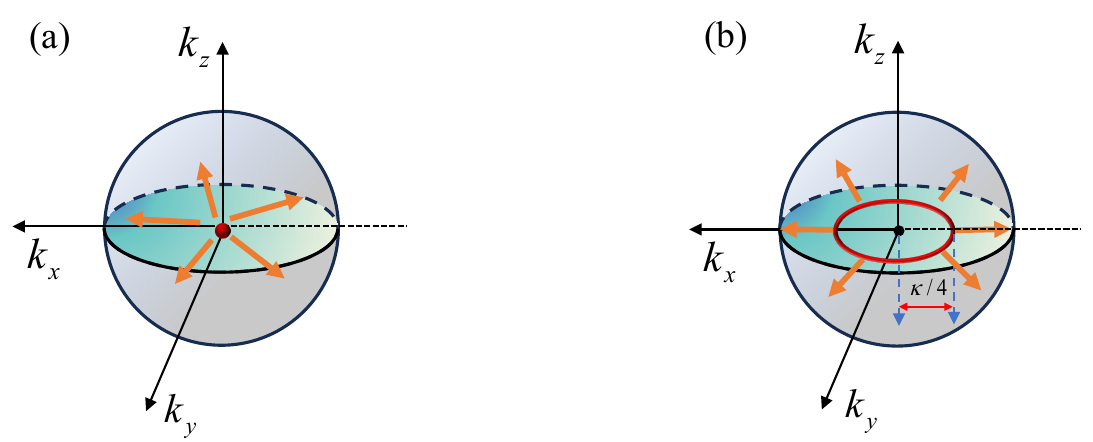}\caption{Construction of the WER. (a) Representation of the Dirac point in
the Hermitian limit. The reciprocal space is defined by $k_{x}=\text{Re}(\Omega)$,
$k_{y}=\text{Im}(\Omega)$, and $k_{z}=\Delta/2$. A degeneracy appears
at the origin, corresponding to a Dirac point carrying a topological
charge of one half. (b) Emergence of the WER in the non-Hermitian regime.
Upon introducing photon loss characterized by decay rate $\kappa$,
the Dirac point is deformed into a ring of second-order exceptional
points in the $k_{x}\lyxmathsym{–}k_{y}$ plane. The resulting WER
has a radius $R_{\text{WER}}=\kappa/4$, centered at the origin, and
signals a transition from point-like to ring-like topological defects
unique to non-Hermitian systems.}
\label{fig:wide-1}
\end{figure*}

Non-Hermitian physics offers a mathematical framework fundamentally
distinct from that of Hermitian theory, enabling EPs to exhibit topological
structures that have no counterpart in Hermitian systems characterized
by quantized topological invariants and Berry phase \citep{34,35,36,37,38,43,44,45,46,47}.
When a control parameter in a non-Hermitian system is extended from
the real domain into the complex domain, pairs of EPs can evolve into
closed loops known as the Weyl exceptional ring (WER). It has been
shown that a WER, formed by second-order EPs, possesses topological
properties markedly different from the Weyl points found in 
Hermitian systems. Specifically, the surface integral of the Berry
curvature over a manifold enclosing the entire ring yields a non-zero,
quantized Chern number, while a loop encircling the ring twice accumulates
a quantized Berry phase \citep{35}. Up to now, the WER has been observed in several experiments, most of which were restricted to classical systems without any quantum effect  \citep{40,41,42,43,44}. In a recent experiment \citep{47}, a quantum WER was implemented with a superconducting qubit coupled to a leaky resonator, whose topological properties were characterized by quantum state tomography. However, a full control of the system Hamiltonian, particularly the phase, remains an experimentally challenging, so that a symmetry needs to be made to identify the WER and to extract the Chern number.


Here, by employing single-photon interferometric network techniques \citep{48,49,50,51},
we overcome the difficulty precise phase control in quantum systems, thereby enabling a complete simulation of the non-Hermitian EP ring in three-dimensional reciprocal space without invoking any additional symmetry assumptions. We successfully simulate
and fully map out the topology of a two-dimensional non-Hermitian
system exhibiting a WER. Through interferometric network, we not only
realize the WER in the entire complex domain but also reconstruct
the system's eigenstates. By extracting the Chern number and Berry
phase over different parameter manifolds, we observe topological critical
phenomena and provide a comprehensive characterization of the WER's
topological properties throughout the complex reciprocal space. All of these pave the way for exploring more complex non-Hermitian systems and for elucidating their exotic topological properties.
\begin{figure*}
\includegraphics[width=2\columnwidth]{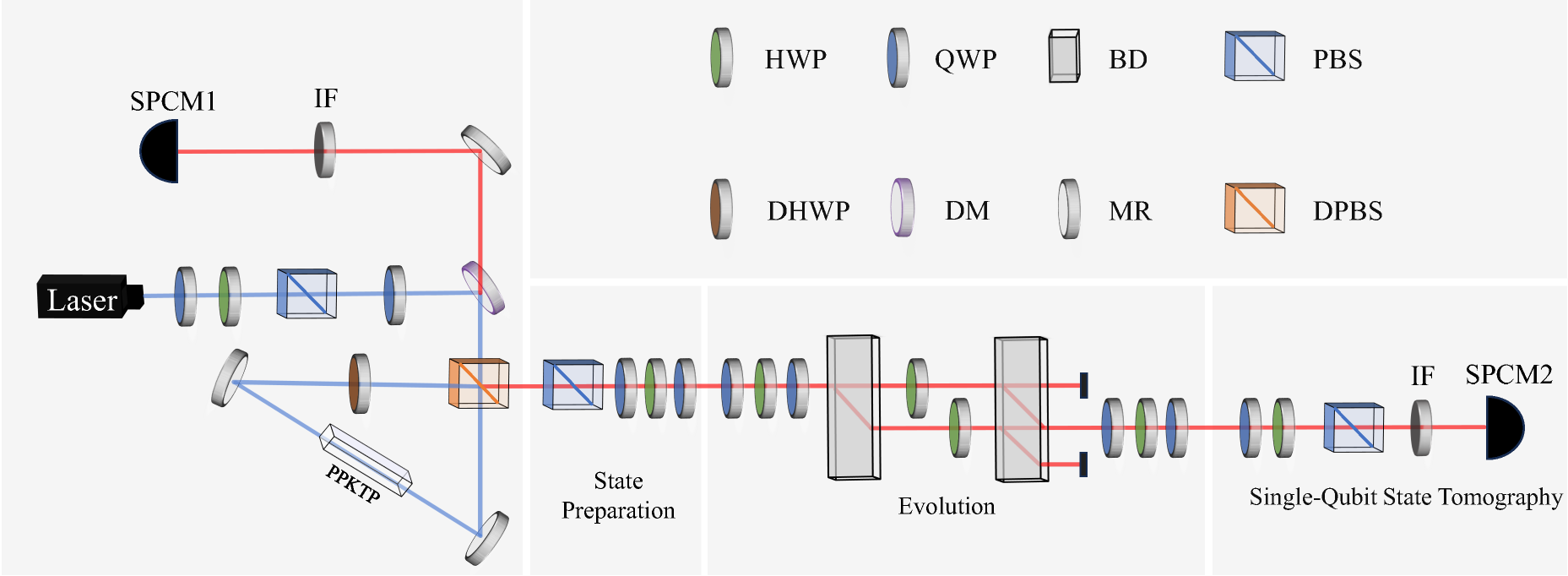}\caption{Experimental setup. A heralded pair of single photons is generated
via type-II spontaneous parametric down-conversion (SPDC) in a periodically
poled potassium titanyl phosphate (PPKTP) crystal. The signal photon
undergoes three stages: (1) State preparation: The photon passes through
a polarizing beam splitter (PBS), two quarter-wave plates (QWPs),
and a half-wave plate (HWP) to initialize the desired polarization
state. (2) Non-unitary evolution: The photon evolves under a simulated
non-Hermitian Hamiltonian using a set of beam displacers (BDs), HWPs,
and QWPs that together implement the time-evolution operator. (3)
Finally, State tomography: The output polarization state is analyzed
via quantum state tomography, using a combination of QWP, HWP, PBS,
and an interference filter (IF), followed by single-photon detection.
The setup enables full reconstruction of the system's evolution in
reciprocal space.}
\label{fig:wide-2}
\end{figure*}

\section{SIMULATION OF EP RING}

In the vicinity of the Dirac point in reciprocal space, as illustrated
in Fig. \ref{fig:wide-1}(a), the system's topological structure
can be described by the following two-component Hamiltonian for a
qubit: 
\begin{equation}
H_{D}=\sum_{\nu=x, y, z}k_{\nu}\sigma_{\nu},
\end{equation}
where $\sigma_{\nu}$ represents Pauli matrices. This Hermitian system
exhibits a degeneracy at the origin, known as a Dirac point, where
two eigenenergies coincide while their corresponding eigenvectors
remain distinct. Such a degeneracy corresponds to a Dirac monopole
carrying a topological charge of two, as illustrated in Fig. \ref{fig:wide-1}(a).
Due to the presence of this topological defect, any parameter manifold
that encloses the degeneracy is topologically distinct from the one
that does not. This topological distinction is characterized by a
topological invariant of the parameter manifold – the Chern number
– which is defined as the integral of the Berry curvature over the
entire reciprocal space.

When dissipation is taken into account, the system's dynamics must
be described using a master equation. However, under the condition
of no quantum jump from the system, the dynamics can instead be effectively
captured by a non-Hermitian (NH) Hamiltonian, given by: 
\begin{equation}
H_{0}=H_{D}-\begin{pmatrix}\frac{1}{2}i\kappa_{0} & 0\\
0 & \frac{1}{2}i\kappa_{1}
\end{pmatrix}\label{eq:2},
\end{equation}
with $\kappa_{0}$ ($\kappa_{1}$) being the decaying rate for the
two basis states $|0\rangle$ and $|1\rangle$ of the qubit, respectively \citep{52,53,54,55}.
In our experiment, the horizontal and vertical polarizations of a
photon serve as the basis states $|0\rangle$ and $|1\rangle$, respectively.
Without loss of generality, we consider the case where dissipation
is present only in the lower component. Under this mapping, the system's
effective non-Hermitian Hamiltonian can be written as: 
\begin{align}
H_{\text{eff}}=\begin{pmatrix}\Delta & \Omega e^{-i\phi}\\
\Omega e^{i\phi}& -\frac{i\kappa}{2}
\end{pmatrix},
\end{align}
here $k_{x}=\text{Re}(\Omega e^{-i\phi})$, $k_{y}=\text{Im}(\Omega e^{-i\phi})$ and $k_{z}=\Delta/2$,
this reformulation is made to be consistent with the notation conventionally
used in experiments. Due to the distinct mathematical structure of
non-Hermitian (NH) Hamiltonians, their eigenvectors must be determined
using the biorthogonal condition, given by $\langle u_{n}^{l}|u_{m}^{r}\rangle=\delta_{m,n}$ \citep{22,24}. Here, $\langle u_{n}^{l}|$ and $|u_{m}^{r}\rangle$ represent the
left and right eigenvectors of the system corresponding to the nth
and mth eigenstates, respectively. Unlike the Hermitian case, the
left and right eigenvectors are not related by Hermitian conjugation.
Instead, they satisfy $H_{\text{eff}}|u_{n}^{r}\rangle=E_{n}|u_{n}^{r}\rangle$
and $\langle u_{n}^{l}|H_{\text{eff}}=\langle u_{n}^{l}|E_{n}$,
where $E_{n}$ is the nth eigenvalue of the system. The two nonorthogonal right eigenvectors
$|u_{n}^{r}\rangle(n=1,2)$ of the effective NH Hamiltonian can be
expressed as:

\begin{equation}
|u_{n}^{r}\rangle=\frac{\Omega^{*}|0\rangle+(E_{n}-\Delta)|1\rangle}{\sqrt{\left|\Omega\right|^{2}+\left|E_{n}-\Delta\right|^{2}}},
\end{equation}
where 
\begin{equation}
E_{\pm}=\frac{2\Delta-i\kappa}{4}\pm\sqrt{\left|\Omega\right|^{2}+\frac{(2\Delta+i\kappa)^{2}}{16}}.
\end{equation}

The non-Hermitian nature of the system transforms the Dirac degeneracy
point in reciprocal space into a ring-shaped structure – known as
a WER – with a radius of $R_{\text{WER}}=\kappa/4$ {[}Fig. \ref{fig:wide-1}(b){]}.
This ring is centered at the origin and lies in the $k_{x}-k_{y}$
plane of reciprocal space. Along the WER, both the eigenenergies and
the eigenvectors coalesce. As the topological defect evolves from
a point-like to a ring structure, the topological properties of the
associated parameter-space manifold now depend not only on its location
but also on its size. This marks a fundamental departure from the
Hermitian case, where topology is solely determined by whether the
degeneracy is enclosed.

\section{EXPERIMENTAL SETUP}

By mapping the spin doublet onto the horizontal and vertical polarizations
of single photon, we simulate the dynamical behavior of the non-Hermitian
Hamiltonian $H_{\text{eff}}$ in reciprocal space. Utilizing a single-photon
interferometric network, we achieve full parameter control across
the entire reciprocal space and extract the systems eigenvectors under
various conditions. This enables a comprehensive characterization
of the topological structure of the WER in a two-dimensional non-Hermitian
system, through which we also observe the emergence of topological
critical phenomena.

As illustrated in the Fig. \ref{fig:wide-2}, our experimental
procedure consists of three main stages: state preparation, non-unitary
evolution, and measurement. In the state preparation stage, we generate
heralded single photons via type-II spontaneous parametric down-conversion
(SPDC) \citep{56}. One photon serves as a trigger, while the other
undergoes the desired dynamical evolution. The heralded photon is
then prepared in the initial state $|H\rangle$ using a polarizing
beam splitter (PBS), two quarter-wave plates (QWPs) and a half-wave
plate (HWP). This state preparation process is depicted in Fig.
\ref{fig:wide-2}. 
\begin{figure*}[t]
	\centering
	\includegraphics[width=\textwidth]{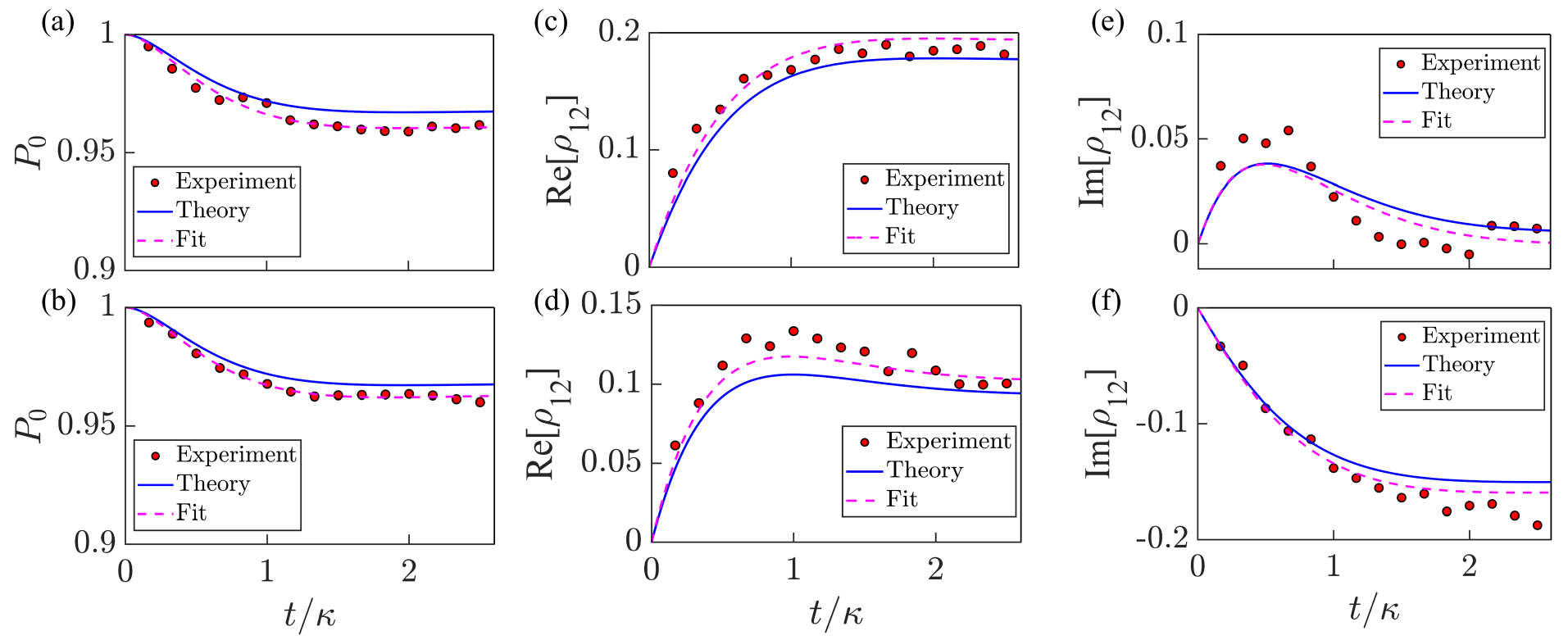}
	\caption{Typical dynamical evolution results. The red dots represent experimental data, the blue solid lines denote theoretical results, and the purple dashed lines correspond to fitting results. The fitting curves are obtained from the experimental data using the method of least squares. (a) and (b) show the time evolution of the normalized density matrix projection onto the basis states $|0\rangle$. (c) and (d) display the real part of the off-diagonal elements of the density matrix, while (e) and (f) show the imaginary part. (a), (c), and (e), and (b), (d), and (f) correspond to system parameters with coupling strengths $\Omega=0.25\kappa$, coupling phases $\phi=\pi/3$ and $2\pi/3$, and detunings $\Delta=0.1\kappa$.
}
	\label{fig:3}
\end{figure*}

\begin{figure*}
\includegraphics[width=2.05\columnwidth]{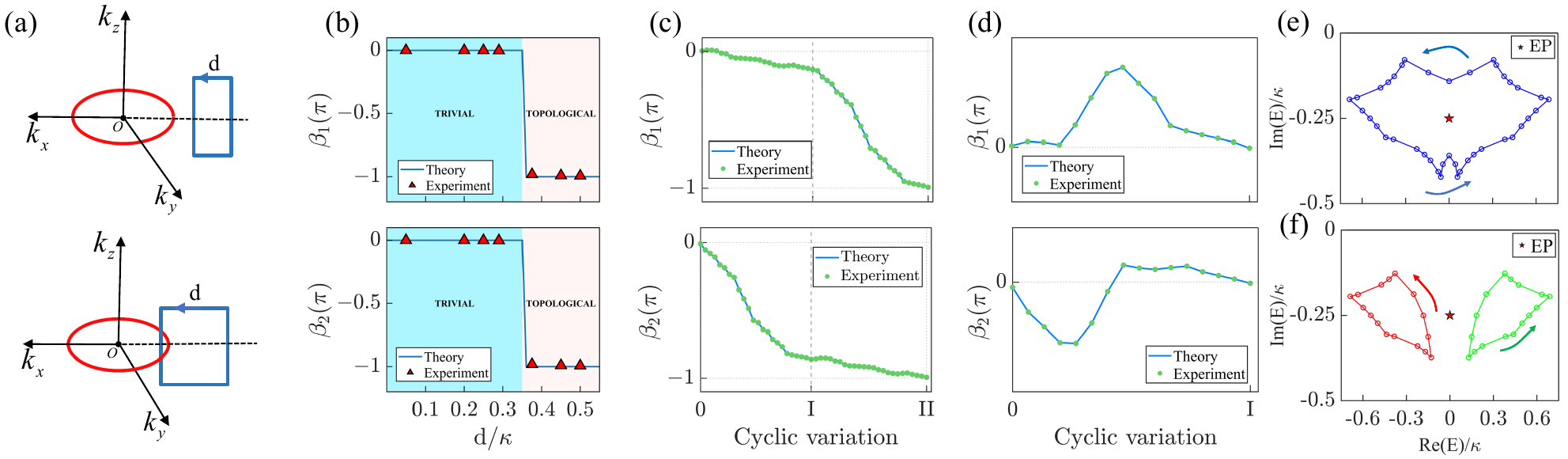}\caption{Topological transition characterized with the Berry phase. (a) Loops
associated with topologically distinct phases. A square-shaped loop
is constructed on the $k_{x}-k_{z}$ plane, with one side fixed at
$k_{x}=\text{Re}(\Omega)=-0.6\kappa$ and a vertical extent of $\kappa/4$.
The loop's side length in $k_x-$direction  $\text{d}$ is varied from $0.35\kappa$ to
$0.85\kappa$ to enclose the WER. (b) Measured Berry phase versus
$\text{d}$. The triangles denote the results associated with the
two pairs of eigenvectors\{$|u_{n}^{r}(\Omega,\Delta)\rangle,|u_{n}^{l}(\Omega,\Delta)\rangle\text{\} with \ensuremath{n=1,2}}$.
The full trajectory of the loop constructed in reciprocal space is
shown in the Appendix C. (c),(d) The corresponding Berry phase after
two cycles ($\text{d}=0.5\kappa$) and one cycle ($\text{d}=0.29\kappa$),
respectively. (e) The spectrum along the summation path of the Berry
phase across two cycles ($\text{d}=0.375\kappa$). Due to the eigenvectors
exchange, the spectrum's summation path $E_{1}$ is the same as $E_{2}$.
(f) The spectrum along the summation path of the Berry phase across
one cycle for $E_{1}$ (left circle) and $E_{2}$ (right circle) ($\text{d}=0.29\kappa$),
respectively. The dots represent the experimental data and the solid
lines denote the theoretical results for the experimental parameters.}
\label{fig:wide-4}
\end{figure*}

Through the interferometric network, the photon undergoes non-unitary
evolution governed by the effective non-Hermitian Hamiltonian $H_{\text{eff}}$,
as shown in Fig. \ref{fig:wide-2}. The corresponding time-evolution
operator $U=e^{-i\hat{H}_{\text{eff}}t}$ for an evolution time $t$
is decomposed into three components: 
\begin{equation}
U=R_{2}(\varphi_{3},\theta_{2},\varphi_{4})L(\theta_{H},\theta_{V})R_{1}(\text{\ensuremath{\varphi_{1},\theta_{1},\varphi_{2}}}).
\end{equation}
Here, each rotation operator $R_{i}(\varphi_{2i-1},\text{\ensuremath{\theta}}_{i},\varphi_{2i})$
$(i=1,2)$ is implemented using a combination of two QWPs set at angles
$\varphi_{i}$ and one HWP set at angle $\text{\ensuremath{\theta}}_{i}$. The polarization-dependent loss operator $L$ is realized using
two beam displacers (BDs) and two HWPs set at angles $\theta_{H}$
and $\theta_{V}$. By tuning the angles of the wave-plate $\{\varphi_{i},\text{\ensuremath{\theta}}_{i},\theta_{H},\theta_{V}\}$ assembly, we achieve precise control over the system parameters, including the phase, as well as the evolution time, thereby enabling accurate quantum simulation of the system’s dynamics (See the appendix
A for details).

Finally, the dynamical evolution of the system is measured via quantum
state tomography, conditioned on coincidence detection between the
signal and herald photons, as shown in Fig. \ref{fig:wide-2}.
Specifically, we measure the probabilities of the photon being projected
onto four polarization states: $\{|H\rangle, |V\rangle, |G_{+}\rangle=(|H\rangle+|V\rangle)/\text{\ensuremath{\sqrt{2}}}, |G_{-}\rangle=(|H\text{\ensuremath{\rangle}}-i|V\rangle)/\text{\ensuremath{\sqrt{2}}}\}$.
These projections are implemented using a combination of a QWP, a
HWP and a PBS. Based on these measurements, we perform a maximum-likelihood
estimation to reconstruct the density matrix and thereby recover the
corresponding dynamical evolution of the system.

By adjusting the angles of the wave plates in the evolution region of Fig. \ref{fig:wide-2}, we achieve precise control of the system parameters, including the phase $\phi$. State tomography then allows us to reconstruct the time evolution of the system’s density matrix. From these results, we extract the dynamical features and obtain the eigenvectors of the system under different parameter settings (See the Appendix B for details). Fig. \ref{fig:3} presents the measured dynamical evolution of the system's density matrix at representative parameter values. Experimental data points are shown as markers, while solid lines represent theoretical simulations. The dashed curve represents the trajectory fitted from the extracted values of $\Delta$ and $\Omega$.
The influence of $\Delta$, $\kappa$, and $\Omega$ on the system dynamics can be directly observed in the populations of the diagonal elements (basis states $|0\rangle$), as illustrated in Fig. \ref{fig:3} (a) and  \ref{fig:3} (b). In contrast, the influence of the phase  $\phi$ on the system dynamics is reflected in the evolution of coherence during the dynamical process, which is clearly demonstrated in the experimental data for the time evolution of the off-diagonal elements of the density matrix, shown in Fig. \ref{fig:3} (c)-(f). These results demonstrate that we have realized precise control over all parameters of the system Hamiltonian, including the phase. This is reflected in the high-fidelity measurements of the density matrix evolution shown in Fig. \ref{fig:3}, which establishes a robust foundation for the subsequent characterization of the system’s topological structure.


\section{OBSERVATION OF BERRY PHASE AND CHERN NUMBER}

The topological characteristics of the WER manifest in two distinct
ways. First, the integral of the Berry curvature over a surface enclosing
the WER yields a nonzero, quantized Chern number. Second,
along a closed trajectory encircling the WER twice, one obtains a
quantized Berry phase. The Berry phase in this context is defined
as \citep{34} 
\begin{equation}
\beta_{n}=i\oint_{2Y}\langle u_{n}^{l}(\Delta,\Omega)|\frac{\partial}{\partial(\Delta,\Omega)}|u_{n}^{r}(\Delta,\Omega)\rangle\cdot dk.
\end{equation}
Here, the path $2Y$ refers to the evolution trajectory in reciprocal space obtained by evolving twice along the blue rectangular loop shown in Fig. \ref{fig:wide-4}(a). This
ensures that the system's eigenvector returns to its original state
upon completing the full trajectory. When the loop encloses the WER,
the acquired Berry phase is quantized at $\pm\pi$; in contrast, for
loops that do not enclose the WER, the Berry phase is zero. By directly
extracting the system's eigenvectors from experimental data, we circumvent
the breakdown of adiabatic following in non-Hermitian systems caused
by their inherent chiral nature. 

For experimental convenience, we choose a rectangular loop lying in
the $k_{x}=\text{Re}(\Omega)-k_{z}=\Delta/2$ plane of reciprocal
space, with one edge fixed at $k_{x}=\text{Re}(\Omega)=-0.6\kappa$
and a height fixed at $\kappa/4$. Whether this loop encloses the
WER depends on the position of the opposite edge of the rectangle,
as illustrated in Fig. \ref{fig:wide-4}(a). Using a discrete summation
approximation, we investigate the relationship between the Berry phase
and the parameters $\Delta$ and $\Omega$, given by:

\begin{equation}
\beta_{n}(\Delta,\Omega)=i\sum_{p=1}^{2m}\langle u_{n,p}^{l}(\text{\ensuremath{\Delta}}\text{,\ensuremath{\Omega}})|u_{n,p+1}^{r}(\text{\ensuremath{\Delta}}\text{,\ensuremath{\Omega}})\rangle-1,
\end{equation}
where, $m$ denotes the number of steps required to traverse each
loop, and $n$ labels the two initial eigenstates. Fig. \ref{fig:wide-4}(b)
shows how the Berry phases $\beta_{1}$ and $\beta_{2}$ evolve as
the position of the opposite edge of the rectangular loop is varied.
As the loop is gradually expanded, we observe that the Berry phases
exhibit sharp transitions near $k_{x}=\text{Re}(\Omega)=-\kappa/4$;
upon crossing this critical point, they rapidly drop to $-\pi$, indicating
the occurrence of a topological phase transition.

Fig. \ref{fig:wide-4}(c) and \ref{fig:wide-4}(d) further illustrate
the step-by-step evolution ($p>1$) of the Berry phases $\beta_{1}$
and $\beta_{2}$ for the two eigenstates under different loop configurations.
In Fig. \ref{fig:wide-4}(c), where the rectangular loop encloses
the WER – corresponding to the opposite edge positioned at $k_{x}=\text{Re}(\Omega)=-\kappa/6$
– the measured Berry phases ($\beta_{1}$ and $\beta_{2}$) in the
topological phase are about $-0.993\pi$. In contrast, Fig. \ref{fig:wide-4}(d)
shows the case where the loop does not enclose the WER, with the opposite
edge at $k_{x}=\text{Re}(\Omega)=-0.31\kappa$; here, the Berry phases
($\beta_{1}$ and $\beta_{2}$) in the trivial phase are measured
to be about $-1.8\times10^{-3}\pi$ and $-1.6\times10^{-3}\pi$, respectively.
It is important to note that in the measurement process, the extracted
eigenvectors inevitably carry arbitrary global phase factors. To eliminate
this ambiguity, we apply a phase correction to the right eigenvector
$|u_{n}^{r},p+1\rangle$ at each intermediate step $p$, using a rotation
angle $\phi_{n}^{p+1}$ = $-\arg\left[\ln\langle u_{n}^{l},_{p+1}|\bar{u}_{n,p}^{r}\rangle\right]$ \citep{47},
\begin{equation}
|\bar{u}_{n,p}^{r}\rangle=e^{i\phi_{n}^{p+1}}|u_{n}^{r},_{p+1}\rangle.
\end{equation}

This procedure allows us to faithfully track the evolution of the
system's Berry phase. To further verify the continuity of the eigenstates
along the chosen path, we present in Fig. \ref{fig:wide-4} the
corresponding evolution of the eigenenergy spectrum throughout the
entire process. The spectrum along the path is shown in Fig. \ref{fig:wide-4}(e), As shown in the figure, in the experiment we label different trajectories by the loop's side length in $k_x-$direction  $\text{d}$.
When $\text{d}>0.35\kappa$, the WER is encircled, and the eigenvectors
exchange occurs near $k_{x}=0$. After the completion of two cycles
{[}Fig. \ref{fig:wide-4}(e){]}, the eigenvector is restored and
obtains a phase difference, approaching $-\pi$, which is the Berry
phase. However, when $\text{d}<0.35\kappa$, the WER is not encircled
and there is no eigenvectors exchange. As a result, the eigenvector
is fully restored in only one complete cycle {[}Fig. \ref{fig:wide-4}(f){]},
and consequently, the corresponding Berry phase drops to $0$. 
\begin{figure*}
\includegraphics[width=2.075\columnwidth]{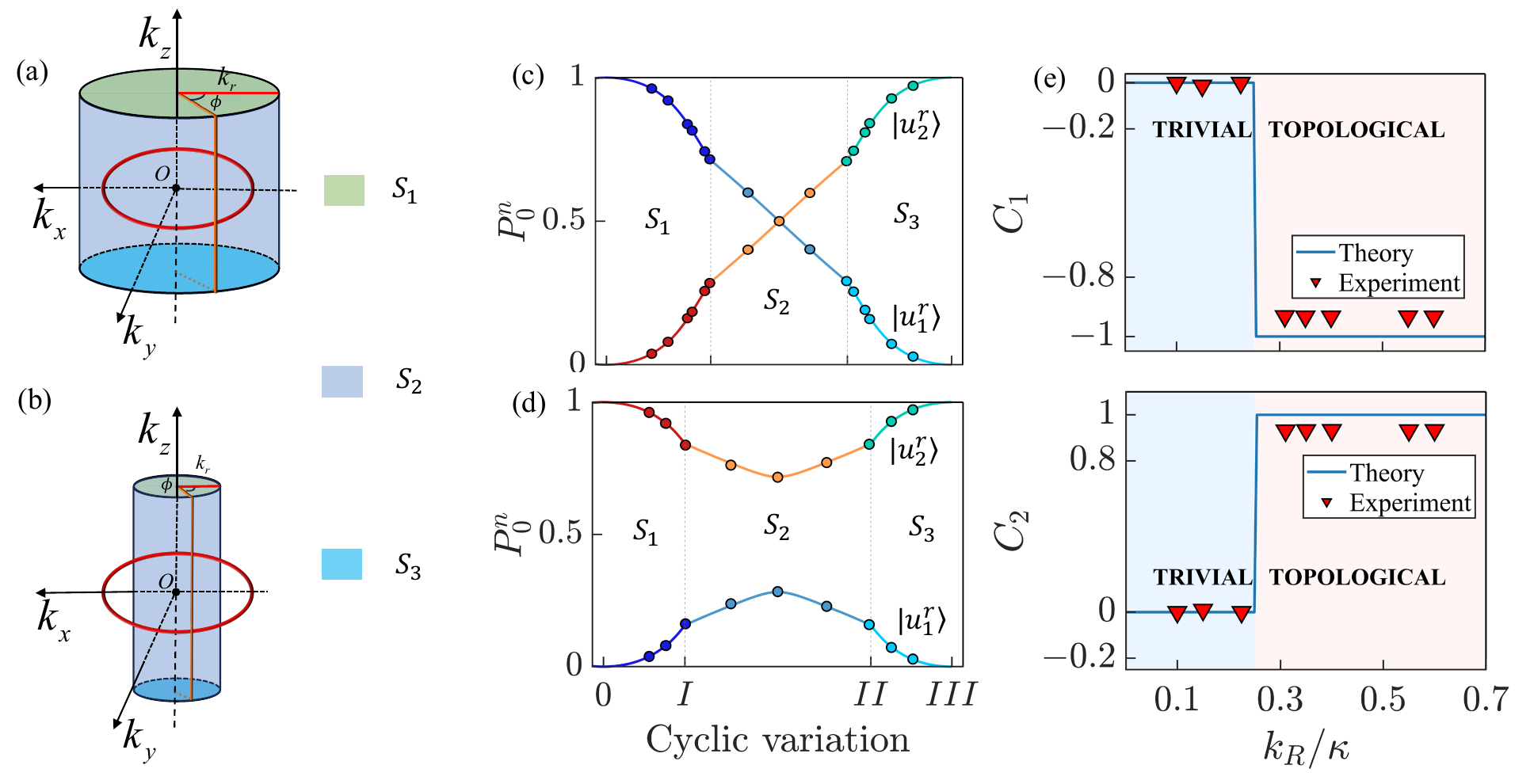}\caption{Topological transition characterized with Chern number. (a),(b) Manifolds
associated with topologically distinct phases. The cylindrical manifolds
are centered at the origin of the reciprocal space. When the WER is
enclosed in the manifold (a), the quantized Berry flux emanated from
the WER pierces through the manifold. When the sphere is located inside
the ring (b), the amount of the Berry flux entering the manifold equals
the amount going out of it. We simulated three trajectories with phase angles $\phi=0$, $\pi/3$, and $2\pi/3$ respectively. 
(c), (d) Measured populations of $\left|0\right\rangle $
in $|u_{r}^{n}\rangle\text{ \ensuremath{(n=1,2)}}$ for $k_{R}=0.35\kappa$
(c) and $0.25\kappa$ (d). Here $(k_{r},\phi,k_z)$ denote the cylindrical
coordinates for the control parameter. The
closed path is divided into three segments. In segment $0\text{–}\ensuremath{\mathrm{I}}$,
the integration is performed along the radial direction $k_{r}\in(0,k_R)$
with fixed $k_{z}=\Delta_0/2$. In segment $\ensuremath{\mathrm{I}}\text{–}\ensuremath{\mathrm{II}}$,
the path proceeds along the $k_{z}$-axis with fixed $k_{x}=k_R$, where
$k_{z}$ varies from $\Delta_{0}/2$ to $-\Delta_{0}/2$. Segment $\mathrm{II}\lyxmathsym{–}\mathrm{III}$
returns along the radial direction with fixed $k_{z}=-\Delta_0/2$,
integrating $k_{r}$ from $k_R$ to $0$. (e) Measured Chern number versus
the radius of the manifold. The triangles denote the results associated
with the two right eigenvectors $|u_{1}^{r}\rangle$ and $|u_{2}^{r}\rangle$,
respectively. The full trajectory of the loop constructed in reciprocal
space is shown in the Appendix C.}
\label{fig:wide-5}
\end{figure*}

As for the other manifestation of the WER's topological properties
– the quantized topological invariant known as the Chern number –
our experiment employs a cylindrical manifold. This manifold is defined
in reciprocal space by three parameters: the radius $k_{r}$ in the
$k_{x}\lyxmathsym{–}k_{y}$ plane, the azimuthal angle (phase) $\phi$ measured
from the positive $k_{x}$-axis, and the height $k_{z}$ along the
$k_{z}=\Delta/2$ direction. The corresponding Berry connection is
defined as:

\begin{equation}
A_{\phi}^{n}=-i\langle u_{n}^{l}(k_{r},\phi,k_z)|\partial_{\phi}|u_{n}^{r}(k_{r},\phi,k_z)\rangle,
\end{equation}
\begin{equation}
A_{z}^{n}=-i\langle u_{n}^{l}(k_{r},\phi,k_z)|\partial_{k_z}|u_{n}^{r}(k_{r},\phi,k_z)\rangle,
\end{equation}
\begin{equation}
A_{k_{r}}^{n}=-i\text{\ensuremath{\langle}}u_{n}^{l}(k_{r},\phi,k_z)|\partial_{k_{r}}|u_{n}^{r}(k_{r},\phi,k_z)\rangle.
\end{equation}

In this case, the expression for the Chern number can be decomposed
into three parts, each corresponding to the contribution by one of
the three edges of the rectangular contour \citep{57,34,58}, the three differently colored regions shown in the Fig. \ref{fig:wide-5}(a) and (b):

\begin{equation}
\begin{aligned}C_{n} & =\frac{1}{2\pi}\int_{0}^{2\pi}d\phi\left[\int_{0}^{k_R}k_rdk_r\frac{\partial}{\partial k_r}A_{\phi}^{n}(k_r,k_z=\Delta_0/2)\right.\\
 & \quad\left.+\int_{\Delta_0/2}^{-\Delta_0/2}dk_z\frac{\partial}{\partial k_z}A_{\phi}^{n}(k_R,k_z)\right.\\
 & \quad\left.+\int_{k_R}^{0}k_rdk_r\frac{\partial}{\partial k_r}A_{\phi}^{n}(k_r,k_z=-\Delta_0/2)\right],
\end{aligned}
\label{C1}
\end{equation}
here, $\Delta_0/2$ corresponds to the $k_z$ values of the top and bottom surfaces of the cylinder in reciprocal space, while $k_R$ represents the radius of the cylindrical side surface in reciprocal space. When $k_R>\kappa/4$, the WER is enclosed by the cylindrical manifold,
as illustrated in Fig. \ref{fig:wide-5}(a). In this case, the
Berry flux emanating from the WER penetrates the manifold, resulting
in a nonzero Chern number of $\pm1$. In contrast, when $k_R$ $<\kappa/4$,
the cylinder lies entirely within the WER, as shown in Fig. \ref{fig:wide-5}(b).
Here, the inward and outward Berry flux contributions cancel each
other, yielding a Chern number of zero. Thanks to the precise controllability of the phase, we do not need to invoke the symmetry assumptions used in earlier experiments \citep{59,60}. In our experiment, instead of performing a continuous integration over the phase $\phi$ as required in Eq.(\ref{C1}), we determine the Chern number of the corresponding manifold by averaging the measurement outcomes obtained at discrete values of $\phi=0$, $\pi/3$, and $2\pi/3$. The cross sections shown in Fig.
\ref{fig:wide-5}(a) and \ref{fig:wide-5}(b) correspond to the measurement planes at different $\phi$ that were selected in our experiment.

To gain deeper insight into the physics of this topological phase
transition, we present the evolution of the eigenvectors in Fig.
\ref{fig:wide-5}(c) and \ref{fig:wide-5}(d), when the WER is enclosed
{[}Fig. \ref{fig:wide-5}(c){]} or not enclosed {[}Fig. \ref{fig:wide-5}(d){]}
by the manifold. For clarity, we represent the eigenvectors by their
projections onto the basis states $(1,0)^{\text{T}}$ denoted $P_{0}$.
The different colors correspond to the three distinct contributions
defined in the decomposition of the Chern number expression of Eq.(\ref{C1}).

In Fig. \ref{fig:wide-5}(c), where the WER is enclosed by the
manifold ($k_{R}=0.35\kappa$), the eigenvectors do not undergo inversion
and the resulting measured Chern number ($C_{1}$ and $C_{2}$) are
about $\pm0.933$. In contrast, Fig. \ref{fig:wide-5}(d) corresponds
to the case where the WER lies outside the manifold ($k_{R}=0.225\kappa$).
Here, the eigenvectors invert upon crossing the $k_{x}$-axis. the
resulting measured Chern number ($C_{1}$ and $C_{2}$) are about
$\pm3.1\times10^{-4}$.

In Fig. \ref{fig:wide-5}(e), we show the evolution of the Chern
number as a function of the manifold radius $k_{r}$. As expected,
the experimental results are in good agreement with theoretical predictions.
When the manifold radius exceeds the critical value $\kappa/4$, the
measured Chern number approaches $\pm1$. As $k_{R}$ is reduced to
the critical value $\kappa/4$, a topological transition occurs, marked
by a sudden drop of the Chern number to near zero. Due to experimental
limitations particularly when $k_r$ ($\Omega$) is small, accurate extraction
of eigenvectors becomes challenging, which is the primary source of
deviation between our measured and theoretical Chern numbers. Additional
data illustrating this evolution are provided in the Appendix C. Notably,
this behavior highlights a key distinction between non-Hermitian systems
and their Hermitian counterparts. For Hermitian systems with point-like
topological defects, such as Dirac points, a topological phase transition
only occurs when the manifold is translated such that it no longer
encloses the defect; it cannot be achieved by simply shrinking a manifold
centered on the Dirac point(See the Appendix D for details).

\section{CONCLUSION }

We have constructed the WER in reciprocal space and revealed its structure
through both Berry phase and Chern number measurements. By employing
a single-photon interferometry network, we achieve precise control
over the system's phase in a fully quantum setting, thereby enabling
a complete characterization of the topological structure and properties
of the Weyl exceptional ring across the entire complex reciprocal space without the need for any additional symmetry assumptions.
In both the Berry phase and the topological Chern number, we observed
clear signatures of topological phase transitions. As the rectangular
loop was continuously contracted, transitioning from enclosing the
WER to excluding it, the Berry phase dropped sharply from $\pm\pi$
to zero, which signaling a topological transition. Similarly, when
the cylindrical manifold centered at the origin in reciprocal space
was reduced in size, the associated Chern number transitioned from
$\pm1$ to zero. These anomalous topological features, absent in Hermitian
systems, arise from the non-Hermitian nature of the system, which
transforms a point-like topological defect in the Hermitian case into
a ring-like defect. Moreover, our experimental approach can be extended
to probe higher-order EP topologies. All of these results lay the groundwork for exploring more complex non-Hermitian systems and for elucidating their exotic topological properties.

\begin{acknowledgments}
We thank Wei Yi and Lei Xiao for valuable discussions. This work was supported by the National Natural Science Foundation of China (12274080, 12474356,
12174058), National Youth Science Foundation of China (12204105) and Educational Research Project for Young and Middle Aged Teachers of Fujian Province (JAT241005).
\end{acknowledgments}

\appendix

\section{Experimental realization of (non)unitary evolution for two-level
system}

In order to simulate the time evolution of these two- level systems,
we encode the basis states in the horizontal and vertical polarizations
of a single photon, with $|H\rangle=(1,0)^{\text{T}}$ and $|V\rangle=(0,1)^{\text{T}}$.
We then extend the recipe in Ref. {\citep{61}} to implement a time
evolution driven by an arbitrary non-Hermitian Hamiltonian in the
two-dimensional Hilbert space.

First, we numerically estimate the time-evolution operator, in general,
is given by a $2\times2$ matrix $T=\begin{pmatrix}t_{11} & t_{12}\\
t_{21} & t_{22}
\end{pmatrix}$ with complex elements $t_{ij}=|t_{ij}|e^{i\phi_{ij}}$, $\phi_{ij} \in \mathbb{R}
$.
To proceed, we define the rotation $(S)$ 
\begin{align}
S(\theta) & =\begin{pmatrix}\cos\theta & \sin\theta\\
-\sin\theta & \cos\theta
\end{pmatrix},
\end{align}
and phase-shift $(P_{\pm})$ operators: 
\begin{equation}
P_{+}=\begin{pmatrix}e^{i\theta} & 0\\
0 & 1
\end{pmatrix},P_{-}=\begin{pmatrix}1 & 0\\
0 & e^{i\theta}
\end{pmatrix}.
\end{equation}
Following Ref. {\citep{61}}, we then decompose the matrix $T$
into a series of rotation and shift operators, as well as a loss operator
$L$, which accounts for the nonunitarity of the time evolution. Specifically
\begin{equation}
\begin{aligned}T & =P_{-}(\phi{}_{21})P_{+}(\phi_{11})S(\text{arctan}\left|\frac{t_{21}}{t_{11}}\right|)P_{-}(\text{arg}t_{22}^{'})\\
 & \quad\times P_{+}(\text{arg}t_{12}^{'})S(\gamma_{1})R_{HWP}(\pi/4)LS(\gamma_{2})P_{+}(-\text{arg}t_{12}^{'}),
\end{aligned}
\label{eq:16}
\end{equation}
where we have 
\begin{align}
t_{11}^{'} & =\left|t_{11}\right|\cos(\text{arg}\left|\frac{t_{21}}{t_{11}}\right|)+\left|t_{21}\right|\sin(\text{arg}\left|\frac{t_{21}}{t_{11}}\right|),\\
t_{12}^{'} & =\left|t_{12}\right|\cos(\text{arg}\left|\frac{t_{21}}{t_{11}}\right|)e^{i(\phi_{21}-\phi_{11})}\nonumber \\
 & \quad+\left|t_{22}\right|\sin(\text{arg}\left|\frac{t_{21}}{t_{11}}\right|)e^{i(\phi_{22}-\phi_{21})},\\
t_{22}^{'} & =-\left|t_{12}\right|\sin(\text{arg}\left|\frac{t_{21}}{t_{11}}\right|)e^{i(\phi_{21}-\phi_{11})}\nonumber \\
 & \quad+\left|t_{22}\right|\cos(\text{arg}\left|\frac{t_{21}}{t_{11}}\right|)e^{i(\phi_{22}-\phi_{21})}.
\end{align}
Here, the loss operator is given by 
\begin{equation}
L=\begin{pmatrix}0 & \mu_{2}\\
\mu_{1} & 0
\end{pmatrix},
\end{equation}
where 
\begin{align}
\mu_{j} & =\sqrt{\frac{\left|\tilde{t}_{11}\right|^{2}+\left|\tilde{t}_{22}\right|^{2}+\left|\tilde{t}_{12}\right|^{2}+(-1)^{j-1}\sqrt{q_{j}^{2}+4p_{j}^{2}}}{2}}\nonumber, \\
p_{j} & =\left|\tilde{t}_{1,3-j}\tilde{t}_{3-j,2}\right|\nonumber, \\
q_{j} & =\left|\tilde{t}_{11}\right|^{2}-\left|\tilde{t}_{22}\right|^{2}+(-1)^{j-1}\left|\tilde{t}_{12}\right|^{2}\nonumber, \\
\tilde{t}_{11} & =\left|t_{11}\right|\cos(\text{arctan}\left|\frac{t_{21}}{t_{11}}\right|)+\left|t_{21}\right|\sin(\text{arctan}\left|\frac{t_{21}}{t_{11}}\right|)\nonumber, \\
\tilde{t}_{12} & =\left|t_{12}\right|\cos(\text{arctan}\left|\frac{t_{21}}{t_{11}}\right|)e^{i(\varphi_{12}-\varphi_{11})}\nonumber \\
 & \quad+\left|t_{22}\right|\sin(\text{arctan\ensuremath{\left|\frac{t_{21}}{t_{11}}\right|}})e^{i(\varphi_{22}-\varphi_{21})}\nonumber ,\\
\tilde{t}_{22} & =-\tilde{t}_{12}\sin(\text{arctan\ensuremath{\left|\frac{t_{21}}{t_{11}}\right|}})e^{i(\varphi_{12}-\varphi_{11})}\nonumber \\
 & \quad+\left|t_{22}\right|\cos(\text{arctan\ensuremath{\left|\frac{t_{21}}{t_{11}}\right|}})e^{i(\varphi_{22}-\varphi_{21})}.
\end{align}
Finally, we group the operators in Eq. (\ref{eq:16}) into three parts,
\begin{equation}
\begin{aligned}T & =P_{-}(\phi{}_{21})P_{+}(\phi_{11})S(\text{arctan}\left|\frac{t_{21}}{t_{11}}\right|)P_{-}(\text{arg}t_{22}^{'})\\
 & \quad\times P_{+}(\text{arg}t_{12}^{'})S(\gamma_{1})R_{HWP}(\pi/4)LS(\gamma_{2})P_{+}(-\text{arg}t_{12}^{'}),
\end{aligned}
\label{eq:22}
\end{equation}
thus reproducing the time-evolution operator in the main text.

Experimentally for each given set of ($\Delta,\Omega$) in Eq. (\ref{eq:2})
of the main text, a set of mutually independent setting angles ($\varphi_{1},\theta_{1},\varphi_{2},\theta_{H},\theta_{V},\varphi_{3},\theta_{2},\varphi_{4}$)
are determined. While the choice of setting angles is not unique (due
to the extra degrees of freedom), our protocol chooses the analytically
available one, according to Ref {\citep{61}}.

As illustrated in Fig. \ref{fig:wide-2}, non-unitary evolution part,
we implement $U$ according to 
\begin{equation}
U=R_{2}(\varphi_{3},\theta_{2},\varphi_{4})L(\theta_{H},\theta_{V})R_{1}(\text{\ensuremath{\varphi_{1},\theta_{1},\varphi_{2}}}),
\end{equation}
where the rotation operator $R_{i}(\varphi_{2i-1},\text{\ensuremath{\theta}}_{i},\varphi_{2i})$
$(i=1,2)$ is realized using a sandwich-type wave-plate set, including
a HWP at the setting angle $\theta_{i}$ and two QWPs at $\varphi_{i}$, respectively. The polarization-dependent loss operator $L=\begin{pmatrix}0 & \sin2\theta_{V}\\
\sin2\theta_{H} & 0
\end{pmatrix}$ is realized by a combination of two BDs and two HWPs with setting
angles $\theta_{H}$ and $\theta_{V}$. The setting angles $\{\theta_{i},\varphi_{i},\theta_{H},\theta_{V}\}$
are fixed according to the numerically calculated $U$. We note that
Eq. (\ref{eq:22}) enables us to implement arbitrary nonunitary operators
for a two-level system with different setting angles.

\section{Fitting Procedure for Extracting Eigenstates}

\begin{figure*}
\includegraphics[width=2\columnwidth]{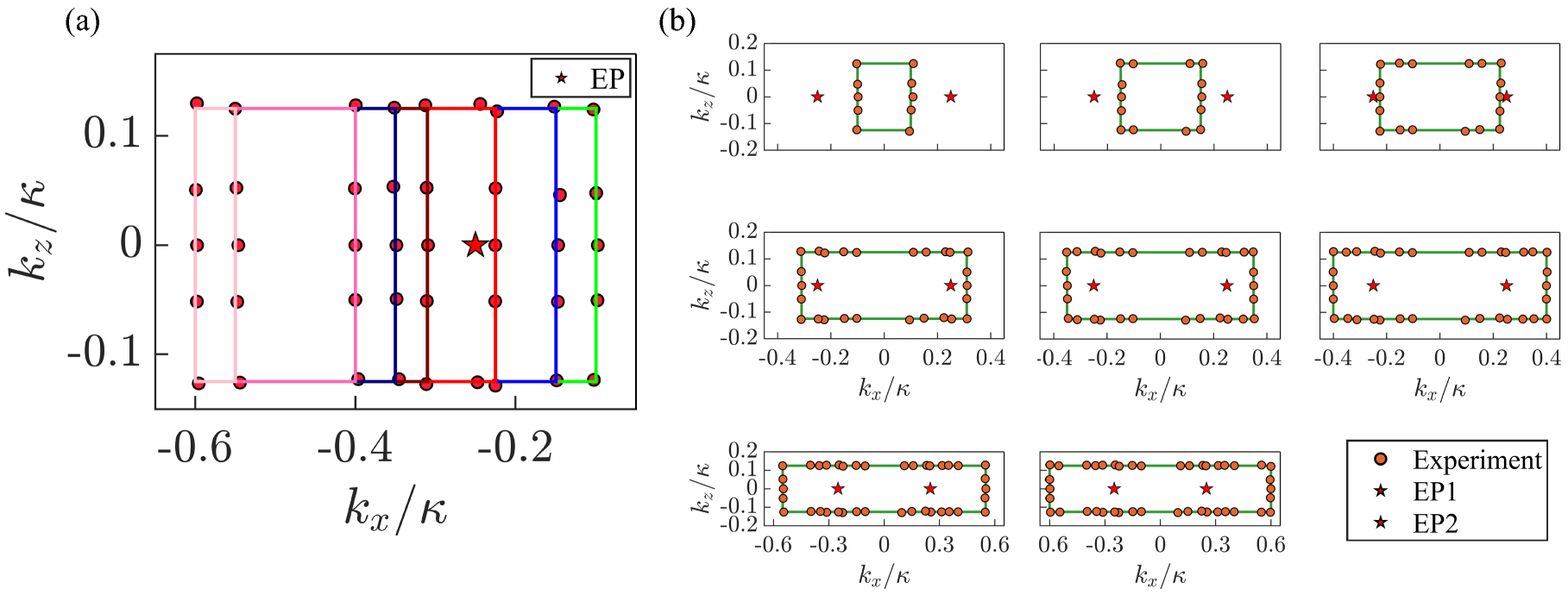}\caption{(a) The summation path of the Berry phase on the $k_{x}-k_{z}$ plane
($k_{y}=0$) for $\text{d}=0.05\kappa,0.2\kappa,0.25\kappa,0.29\kappa,0.375\kappa,0.45\kappa,0.5\kappa$.
The red star marks the EP. The lines denote the theoretical path and
the red dots represent the experimental parameters. (b) The summation
path of the Berry phase on the $k_{x}-k_{z}$ plane ($k_{y}=0$).
The spectrum shifts from left to right for $k_{R}=0.1\kappa,0.15\kappa,0.225\kappa,0.31\kappa,0.35\kappa,0.4\kappa,0.55\kappa,0.6\kappa$.
The lines denote the theoretical path and the orange dots represent
the experimental parameters.}
\label{fig:wide-6}
\end{figure*}
To quantitatively extract the eigenvectors and corresponding observables
from the experimental data, we employed a least-squares fitting method.
This procedure enables us to estimate the parameters $\Delta$ and
$\Omega$ for each point in reciprocal space by fitting the measured
polarization states to theoretical predictions governed by the effective
non-Hermitian Hamiltonian, it should be noted here that 
$\Omega$ is a complex quantity.

The polarization state of the single photon after evolution is characterized
by its Stokes parameters($S_{1},S_{2},S_{3}$), which are reconstructed
via quantum state tomography. Theoretically, these parameters are
functions of the system parameters $\Delta$, $\Omega$, and $\kappa$, through the evolved density matrix $\rho(t;\Delta,\Omega)$. For
each set of experimental measurements \{$S_{i}^{exp}$\}, we define
a cost function: 
\begin{align}
\chi^{2}(\Delta,\Omega) & =\sum_{i=1}^{3}[S_{i}^{exp}-S_{i}^{th}(\Delta,\Omega)]^{2},
\end{align}

where $S_{i}^{th}$ are the theoretically calculated Stokes parameters.
The optimal fitting parameters are obtained by minimizing this cost
function using a nonlinear least-squares algorithm. This approach
allows us to reconstruct the system's eigenvectors $|u_{n}^{r}\rangle$
and $\langle u_{n}^{l}|$ with high fidelity over the entire complex
parameter space. The extracted eigenstates are subsequently used to
compute the Berry phase and Chern number along specified loops or
manifolds.

\section{Raw Experimental Data }

We present here the full set of experimental data points used in the
extraction of both the Berry phase and the Chern number.

Fig. \ref{fig:wide-6}(a) shows the set of parameter points used
to calculate the Berry phase, where all summation paths lie in the
$k_{x}-k_{z}$ plane with $k_{y}=0$. The rectangular loops of varying
sizes correspond to different values of loop size $\text{d}=0.05\kappa, 0.2\kappa, 0.25\kappa, 0.29\kappa, 0.375\kappa, 0.45\kappa, 0.5\kappa$
The red star marks the location of the EP. Solid lines represent the
theoretical loop trajectories, while red dots indicate the experimentally
implemented parameters.

Fig. \ref{fig:wide-6}(b) shows the cylindrical summation paths
used to compute the Chern number. The cylindrical surfaces are centered
around the origin and lie in the same $k_{x}-k_{z}$ plane with fixed
$k_{y}=0$. The measured spectral data are collected for cylindrical
radii $k_{R}=0.1\kappa, 0.15\kappa, 0.225\kappa, 0.31\kappa, 0.35\kappa, 0.4\kappa, 0.55\kappa, 0.6\kappa$.
Orange dots indicate the experimental points, and the solid curves
denote the theoretical predictions.

These data form the foundation for the numerical integration used
in the main text to extract Berry phases and Chern numbers from the
experimentally reconstructed eigenvectors.

\section{Calculation of Berry Phase and Chern Number}

The effective non-Hermitian Hamiltonian used in our system is given
by:

\begin{align}
H_{\text{eff}} & =\begin{pmatrix}\Delta & \Omega^{\ast}\\
\Omega & -\frac{i\kappa}{2}
\end{pmatrix}=\begin{pmatrix}\frac{\Delta}{2}+\frac{i\kappa}{4} & \Omega^{*}\\
\Omega & -\frac{\Delta}{2}-\frac{i\kappa}{4}
\end{pmatrix}.
\end{align}
Its eigenvalues are: $E_{1,2}=\frac{2\Delta-i\kappa}{4}\pm\sqrt{\left|\Omega\right|^{2}+\frac{(2\Delta+i\kappa)^{2}}{16}}$,
so $E_{1,2}=\varepsilon\pm\sqrt{\left|\Omega\right|^{2}+\frac{(2\Delta+i\kappa)^{2}}{16}}=\varepsilon\pm\Omega\sqrt{B^{2}+1}$.
The EP occurs when: $i=\pm1$ or $\Delta=0,\left|\Omega\right|=\frac{\kappa}{4},$
$B=\frac{2\Delta+i\kappa}{4\Omega}$. In the analysis of the eigenstates,
the global energy shift $\varepsilon$ does not affect the eigenvectors.
$H_{\text{eff}}=H_{0}+H=\begin{pmatrix}\frac{\Delta}{2}+\frac{i\kappa}{4} & \Omega^{*}\\
\Omega & -\frac{\Delta}{2}-\frac{i\kappa}{4}
\end{pmatrix}$, the corresponding eigenvalues are $E_{\pm}=\pm\Omega\sqrt{B^{2}+1}$.
To determine the eigenstates of the Hamiltonian, we solve the eigenvalue
equation $H_{\text{eff}}|u_{n}^{r}\rangle=E_{\pm}|u_{n}^{r}\rangle$
\begin{alignat}{3}
\begin{pmatrix}\frac{\Delta}{2}+\frac{i\kappa}{4} & \Omega^{*}\\
\Omega & -\frac{\Delta}{2}-\frac{i\kappa}{4}
\end{pmatrix} & \begin{pmatrix}c_{1}\\
c_{2}
\end{pmatrix} & = & E_{\pm} & \begin{pmatrix}c_{1}\\
c_{2}
\end{pmatrix},
\end{alignat}
$\frac{c_{2}}{c_{1}}=\sqrt{B^{2}+1}-B$, we define $\theta$ such
that $\tan\theta=\sqrt{B^{2}+1}-B,\left|u_{1}^{r}\right\rangle =\begin{pmatrix}\cos\theta\\
\sin\theta
\end{pmatrix}$. Analogously, the second eigenvector takes the form: $\left|u_{2}^{r}\right\rangle =\begin{pmatrix}-\sin\theta\\
\cos\theta
\end{pmatrix}$.

When the system parameters adiabatically evolve along a closed path
in reciprocal space enclosing an EP, the eigenvalues continuously
exchange: $E1\leftrightarrow E2$, expressing the eigenvalues as:
$E_{1,2}=\varepsilon\pm\Omega\sqrt{B^{2}+1}=\varepsilon\pm\Omega\sqrt{B+i}\sqrt{B-i}$,
this exchange corresponds to changing the sign of one branch of the
square root. Correspondingly, the eigenstate parameter transforms
as, $\theta\rightarrow-\sqrt{B^{2}+1}-B=-\cot\theta$. The resulting
state transformation after encircling the EP is: $|u_{1}^{r}\rangle\rightarrow-|u_{2}^{r}\rangle,|u_{2}^{r}\rangle\rightarrow|u_{1}^{r}\rangle$.
If the adiabatic path encircles the EP starting from the first eigenstate
perspective: $|u_{1}^{r}\rangle\rightarrow-|u_{2}^{r}\rangle\rightarrow-|u_{1}^{r}\rangle$,
EP: $\Delta=0,\Omega=\pm\kappa/4$. When evolving in reciprocal space
along a closed loop that returns to the starting point, if the path
encloses an EP, the accumulated geometric phase is $\pi$.

The process of calculating the Chern number is analogous to this.
If the integration path lies inside the WER, then when the path crosses
the $k_{x}$-axis, the eigenstate transforms as $|u_{1}^{r}\rangle\rightarrow|u_{2}^{r}\rangle$.
If the path is outside the WER, no such transformation occurs. Consequently,
the Chern number is zero when the integration path is inside the WER,
while it takes values of $\pm1$ when the path encloses the WER. 

\bibliographystyle{apsrev4-2}
\bibliography{citation}

\end{document}